\documentclass[letterpaper]{article} 
\usepackage[preprint]{aaai2027} 
\usepackage[hyphens]{url}  
\usepackage{graphicx} 
\def\UrlFont{\rm}  
\usepackage{natbib}  
\usepackage{caption} 
\usepackage{algorithm}
\usepackage{algpseudocode}

\usepackage{soul, xcolor}

\usepackage{multirow}
\newcommand{\sysname}{PG-TTRL}
\usepackage{newfloat}
\usepackage{listings}
\DeclareCaptionStyle{ruled}{labelfont=normalfont,labelsep=colon,strut=off} 
\floatstyle{ruled}
\newfloat{listing}{tb}{lst}{}
\floatname{listing}{Listing}

\usepackage{booktabs}

\usepackage{amsmath}
\usepackage{amssymb}

\title{Listen Then Reason: Perception-Grounded Test-Time Reinforcement Learning for Large Audio-Language Models}
\author{
    Jiaheng Dong\textsuperscript{\rm 1}\equalcontrib,
    Xiaofeng Yu\textsuperscript{\rm 2}\equalcontrib,
    Jean Honorio\textsuperscript{\rm 1,\rm 4},
    Abhirup Ghosh\textsuperscript{\rm 3},
    Hong Jia\textsuperscript{\rm 2},
    Ting Dang\textsuperscript{\rm 1}
}
\affiliations{

    \textsuperscript{\rm 1}The University of Melbourne, Australia\\
    \textsuperscript{\rm 2}University of Auckland, New Zealand\\
    \textsuperscript{\rm 3}University of Birmingham, United Kingdom\\
    \textsuperscript{\rm 4}ARC OPTIMA, Australia\\
    jiahengd@student.unimelb.edu.au, xyu724@aucklanduni.ac.nz
}

\begin{document}

\maketitle

\begin{abstract}
Large audio-language models (LALMs) are increasingly used for a broader range of audio reasoning tasks. These models typically incorporate audio representations into a large language model (LLM) backbone to enable multimodal reasoning. Recent test-time reinforcement learning (TTRL) methods further improve LLM reasoning capability by leveraging unlabelled test data after pre-training. However, the importance of the perceptual capability of LALMs remains underexplored, particularly how much acoustic evidence is integrated and relied upon during reasoning, and how this contributes to final task performance. This gap limits the development of effective post-training methods like TTRL for audio reasoning. In this work, we first analyse how audio information is integrated and utilised during reasoning process. We quantify layer-wise perceptual reliance and show that stronger acoustic reliance is associated with higher accuracy and a larger performance gain attributable to the audio input. Building on this, we propose Perception-Grounded TTRL (PG-TTRL), which aligns label-free test-time optimisation with perceptually grounded reasoning, encouraging the model to structure its reasoning more strongly on the audio input. Experiments across LALMs and benchmarks show that PG-TTRL consistently improves reasoning performance over both the base models and standard TTRL, showing the value of perceptual-grounding optimisation for test-time audio reasoning. 
\end{abstract}



\vspace{-2pt}
\vspace{-5pt}
\section{Introduction}
Large Audio–Language Models (LALMs) have demonstrated strong reasoning performance across a diverse range of tasks, from spoken question answering to audio-grounded inference~\cite{ma2026mmar,kumar2026mmau,sakshi2025mmau}. These models typically couple a dedicated audio encoder for acoustic perception with a large language model (LLM) backbone for reasoning, enabling end-to-end multimodal understanding. Building on this foundation, post-training techniques, most notably reinforcement learning (RL), have further validated that reasoning capabilities can be meaningfully enhanced beyond pre-training~\cite{guo2025deepseek,shao2024deepseekmath,wen2025sari,diao2025soundmind}. More recently, test-time reinforcement learning (TTRL), which operates without ground-truth labels, has emerged as a promising direction for leveraging unlabeled data at inference and has shown encouraging gains in model performance~\cite{zuo2026ttrl,ttrl2,ttrl3}.

Despite this rapid progress, a growing body of work has begun to scrutinize the internal mechanics of multi-modal large language models. Studies on large vision–language models (LVLMs) in particular have revealed that these models disproportionately rely on their text-based LLM backbone, with the multi-modal perception component specifically visual understanding, contributing far less than expected to final task performance~\cite{asadi2026mirage,lee2025vlind}.  This imbalance may stem from the 
cross-modal alignment training, which is typically optimized for response correctness rather than perceptual fidelity, allowing the model to exploit linguistic shortcuts~\cite{chen2026language}. LALMs inherit this same architectural blueprint and training recipe, with a frozen or lightly fine-tuned audio encoder aligned to a pre-trained LLM, 
and are therefore likely subject to the same perceptual under-utilization. Yet whether, and to what degree, acoustic perception genuinely drives reasoning in LALMs remains an open and largely unexamined question.

This gap carries significant practical consequences, particularly for post-training and test-time updating. Existing RL-based post-training methods for multimodal LLMs are designed around outcome-level reward signals derived from final answer correctness~\cite{ding2025kimi,huang2025vision}. Such rewards are agnostic to how the model arrived at its answer: a correct response driven by language priors is rewarded identically to one grounded in genuine acoustic perception. As a result, RL fine-tuning may reinforce language-side shortcuts rather than improving perceptual grounding, yielding models that appear stronger on benchmarks yet remain brittle when acoustic understanding is genuinely required. This problem is further compounded in TTRL settings, where the absence of ground-truth labels removes even the weak corrective signal that supervised objectives provide, leaving the optimization entirely susceptible to modality-agnostic reward hacking. Without an explicit mechanism to assess and reward perceptual contribution, test-time updating risk optimizing the wrong component of the model, a fundamental misalignment between the learning objective and the true multi-modal bottleneck.

To address these limitations, we first analyze the perceptual contribution of acoustic inputs in LALM reasoning, quantifying how perception quality affects reasoning outcomes. Building on these insights, we propose a \sysname{}, which augment GRPO (Group Relative Policy Optimization) with audio‑perception reliability for test-time updating
. Our experiments demonstrate that \sysname{} improves reasoning accuracy by up to 4.6 points on MMAR and 4.9 points on MMAU, while achieving the strongest Pass@$k$ performance across nearly all sampling budgets. 
Our contributions are summarised below:
\begin{itemize}
    \item We quantify perceptual reliance in LALMs, revealing the perception evolving pattern in reasoning and establishing its relationship with task performance. 
    \item We propose \sysname{}, the first TTRL that incorporates the reliability of audio perceptual grounding into label-free policy optimisation for audio reasoning.
    \item We demonstrate across two LALMs and two audio reasoning benchmarks that PG-TTRL consistently improves reasoning accuracy and achieves stronger performance especially under limited sampling budgets.
\end{itemize}

\section{Related Work}

\subsection{Large Audio--Language Models}
LALMs extend text-based LLMs with the ability to process speech, environmental sounds, and music. Representative systems include Qwen-Audio and Qwen2-Audio, Kimi-Audio, and the Audio Flamingo series~\cite{chu2023qwen,chu2024qwen2,ding2025kimi,goel2025audio}. Despite differences in scale, most LALMs follow a similar architecture: one or more specialised audio encoders extract acoustic representations, a projection or adaptor module maps them into the LLM embedding space, and a pretrained decoder-only LLM performs instruction following, reasoning, and response generation. Training typically combines audio--text alignment or pretraining with multimodal instruction tuning, enabling a shared model to address tasks such as speech recognition, spoken question answering, audio captioning, emotion understanding, and general audio-based inference. Recent benchmarks such as MMAU and MMAR have further expanded evaluation from conventional audio understanding to real-world and multi-step reasoning across speech, sound, and music~\cite{sakshi2025mmau,ma2026mmar}. However, their results also show that current LALMs continue to struggle on tasks requiring both accurate acoustic perception and complex downstream reasoning.

\subsection{Reasoning in LALMs} 
Recent studies strengthen LALM reasoning through chain-of-thought (CoT) supervision and reinforcement-learning-based post-training. Audio-Reasoner constructs the large-scale CoT corpus and fine-tunes Qwen2-Audio on structured, long-form reasoning traces, demonstrating substantial gains across audio understanding and reasoning benchmarks~\cite{zhifei2025audio}. Reinforcement learning has subsequently been introduced to further improve audio reasoning. SARI combines structured CoT supervision with curriculum-guided GRPO, progressively exposing the model to more difficult audio questions and optimising answer accuracy and output structure~\cite{wen2025sari}. SoundMind-RL applies rule-based RL to audio logical reasoning, using rewards for factual correctness, response format, and reasoning completeness across audio-to-text and audio-to-audio settings~\cite{diao2025soundmind}. Collectively, these methods demonstrate the effectiveness of RL and explicit reasoning supervision for enhancing LALM performance. However, their reliance on ground-truth supervision makes them inapplicable to label-free test-time settings.
Moreover, their rewards are primarily outcome-oriented: correctness determines whether the final answer is reinforced, while format or reasoning-length constraints regulate how the response is expressed. Consequently, models may be optimised blindly toward desirable outputs without acquiring the intended reasoning capability, as these rewards do not assess whether a correct trajectory genuinely uses acoustic evidence or instead relies on language priors and textual shortcuts. 

\subsection{Test-Time Reinforcement Learning}
Test-time reinforcement learning (TTRL) updates models directly on unlabelled test instances. TTRL derives pseudo-labels through majority voting over sampled trajectories and optimises the resulting self-generated rewards using GRPO~\cite{zuo2026ttrl}. Subsequent GRPO-based methods improve the reliability and diversity of this process through entropy-aware exploration, fine-grained pseudo-reward estimation, and minority-preserving mechanisms~\cite{ttrl2,ttrl3}. More recently, TTRV~\cite{singh2025ttrv} extends this paradigm to VLMs by constructing frequency- and diversity-based GRPO rewards from repeated multimodal predictions. Despite these developments, existing TTRL methods determine update direction primarily from agreement among model outputs. This is particularly problematic for LALMs because, without ground-truth labels, majority agreement may reflect shared language-side biases and shortcuts across trajectories. A correctness-based reward will reinforce these shortcuts, driving the policy further from acoustic grounding and amplifying perceptual under-utilisation.



\begin{figure*}[t]
    \centering
    \includegraphics[width=\textwidth]{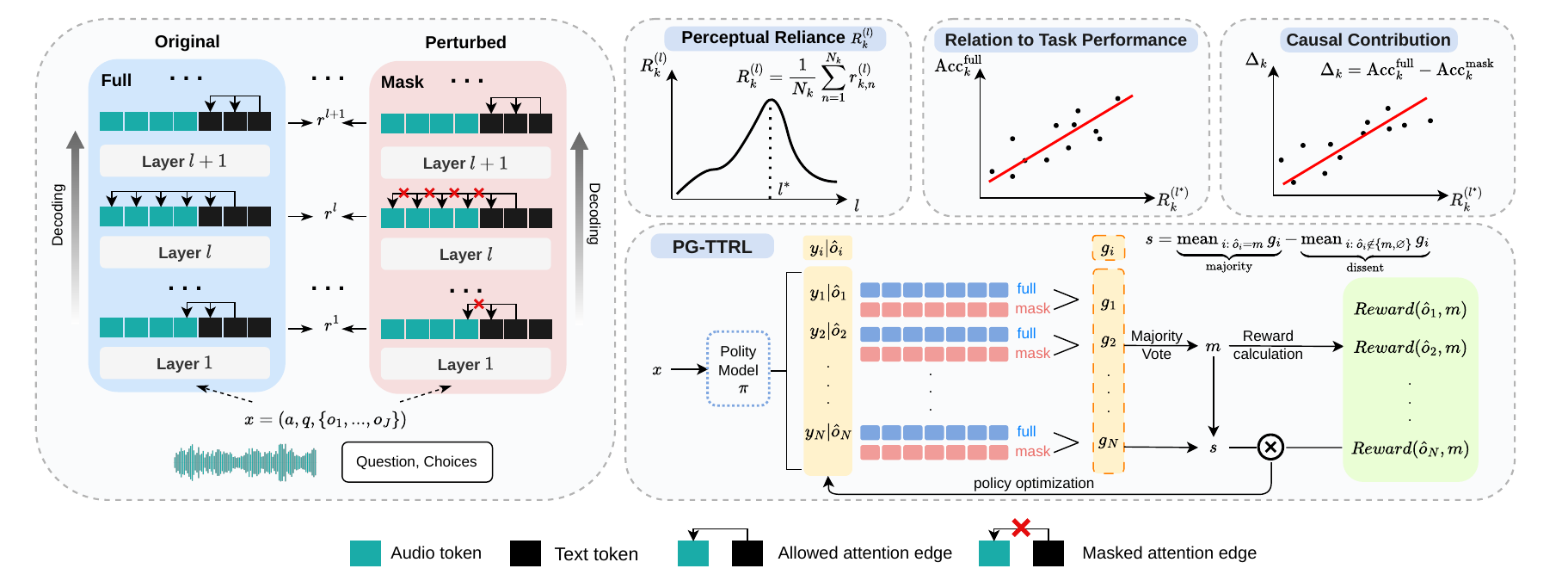}
    \caption{\textbf{Framework Overview.} We compare the original and perturbed conditions to compute the instance-level layerwise perceptual reliance. We then 
    obtain the task-level reliance, and examine its relationship with task accuracy and the causal contribution of acoustic evidence. PG-TTRL computes a trajectory-level grounding score $g_i$ 
    , derives the acoustic grounding margin $s$
    , and uses it to calibrate the advantage for test-time policy optimisation.}
    \vspace{-10pt}
    \label{fig:framework_overview}
\end{figure*}
\section{Perception Grounding in Audio Reasoning}
\label{sec:Methodology-perceptual-analysis}
A LALM processes an audio-grounded query by first encoding the audio input into acoustic representations, which are then projected into the LLM's token space and processed jointly with text tokens. As the joint sequence propagates through the LLM backbone, the model reasons over both modalities, progressively integrating acoustic and linguistic evidence before drawing on this integrated representation for high-level reasoning and answer formation. 
However, it remains unclear how acoustic evidence evolves from cross-modal integration to downstream reasoning, whether the degree of perceptual integration meaningfully predicts task performance, and whether acoustic evidence makes a functional contribution to that performance rather than being bypassed by language-side reasoning.

Therefore, we structure our analysis around three research questions. \textbf{RQ1}: How does acoustic evidence evolve across the LALM backbone for reasoning? 
\textbf{RQ2}: Does stronger 
perceptual grounding predict better performance across audio reasoning tasks? \textbf{RQ3}: Does the integrated acoustic evidence directly contribute to task performance, or can models achieve strong results through language-side reasoning that largely bypasses the audio input? 

\subsection{Audio Perception in Reasoning}
\label{sec:emphasis_boundary_1} 

To address \textbf{RQ1}, we conduct a layer-wise representational analysis of the LALM backbone, tracing how the influence of acoustic evidence evolves across layers and identifying where cross-modal integration is most pronounced and where reasoning over the integrated evidence begins to dominate, as shown in Figure~\ref{fig:framework_overview} left. 

\paragraph{Layer-wise Perceptual Reliance. }
For each layer, we compare the hidden representations under two conditions: (1) the original audio input and (2) a perturbed condition where textual tokens cannot attend to audio tokens. Specifically, given audio tokens $\mathcal{A}$ and text tokens $\mathcal{T}$, we block cross-modal attention by setting $M_{ij}=-\infty$ for $i\in\mathcal{T}$ and $j\in\mathcal{A}$ before the softmax operation. This removes the direct influence of audio, leaving the reasoning dependent on text tokens only. 

Let $\mathbf{h}^{(l)}_{i,\mathrm{full}}$ and $\mathbf{h}^{(l)}_{i,\mathrm{mask}}\in\mathbb{R}^{d}$ denote the hidden representation of text token $i$ at layer $l$ under the original and masked conditions, respectively. We compute representations only for the question and answer-option tokens, denoted by $\mathcal{S}\subseteq\mathcal{T}$, as these capture the task-relevant reasoning process. The layer representation is obtained by mean pooling:
\begin{equation}
\bar{\mathbf{h}}^{(l)}_{c}
=
\frac{1}{|\mathcal{S}|}
\sum_{i\in\mathcal{S}}
\mathbf{h}^{(l)}_{i,c},
\qquad
c\in\{\mathrm{full},\mathrm{mask}\}.
\end{equation}
We define the perceptual reliance $r_{k,n}^{(l)}$ for an instance $n$ in task $\mathcal{D}_k$ as the cosine distance 
between the two pooled representations  $ \bar{\mathbf{h}}^{(l)}_{k,n,\mathrm{full}}$ and $\bar{\mathbf{h}}^{(l)}_{k,n,\mathrm{mask}}$,
where larger values indicate the layer is more strongly influenced by the audio input. Averaging over all instances in task $\mathcal{D}_k$ gives the task-level perceptual reliance:
\begin{equation}
R_k^{(l)}
=
\frac{1}{N_k}
\sum_{n=1}^{N_k}
r_{k,n}^{(l)},
\end{equation}
where $N_k$ is the number of evaluation instances. 

\begin{figure*}[t]
    \centering
    \includegraphics[width=\textwidth]{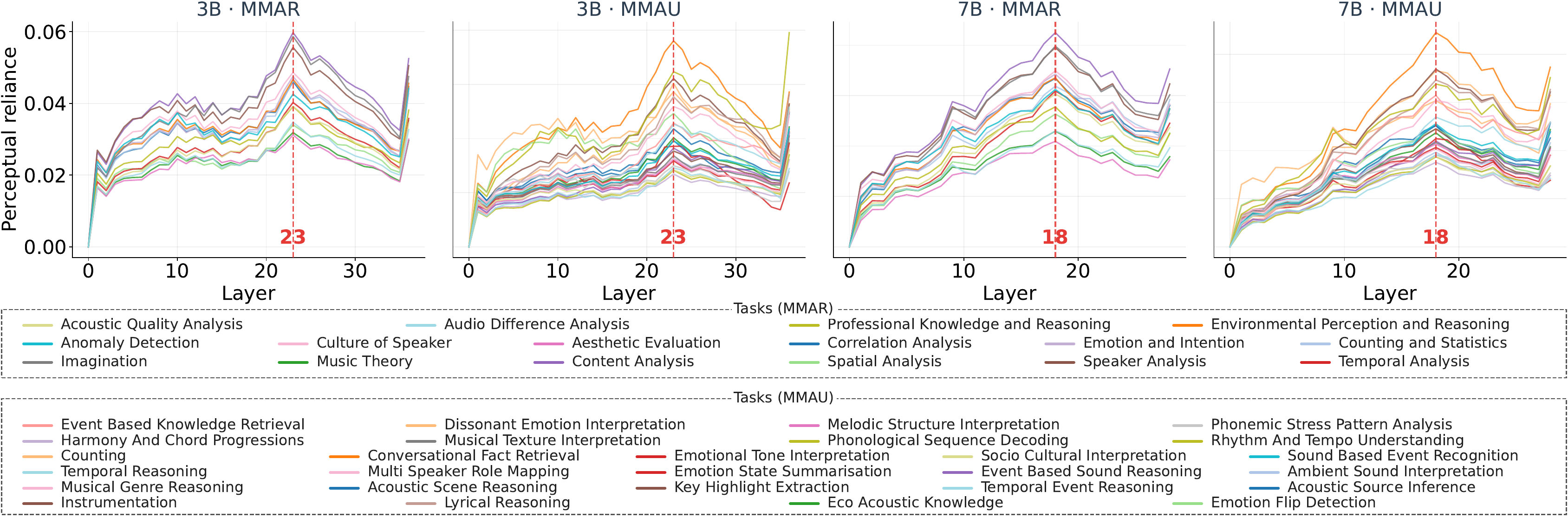}
    \vspace{-20pt}
    \caption{Layer-wise perceptual reliance across MMAR and MMAU for Qwen2.5-Omni-3B and Qwen2.5-Omni-7B.}
    \vspace{-5pt}
    \label{fig:perceptual_reliance_line}
\end{figure*}

\paragraph{Findings. }
Figure~\ref{fig:perceptual_reliance_line} shows a consistent pattern across tasks and model scales. Perceptual reliance increases in the early layers, reaches a maximum at an intermediate layer, and then gradually decreases toward the output layers across all different tasks, model backbones, and datasets. This suggests that early layers progressively emphasise integrating acoustic information, whereas later layers rely increasingly on the integrated representation to emphasise reasoning and answer formation. This pattern is an intrinsic property of current LALMs rather than a model- or data-specific phenomenon. 

We therefore define the perception--reasoning emphasis boundary as the layer with maximum perceptual reliance,
\begin{equation}
l^*
=
\arg\max_{l} R_k^{(l)}.
\end{equation}
which represents the transition from perception-dominated processing to reasoning-dominated computation. We use the peak perceptual reliance, $R_k^{(l^*)}$, as the task-level measure of acoustic dependence in the remainder of the paper.


\subsection{Perceptual Reliance and Task Performance}
To answer \textbf{RQ2}, we investigate whether 
capturing stronger acoustic information leads to improved performance. Specifically, we examine the relationship between perceptual reliance and task performance at the task-level. 

For each task $\mathcal{D}_k$, we measure the task-level 
perceptual reliance
$R_k^{(l^*)}$ at the perception-reasoning emphasize boundary 
and the corresponding task accuracy
$\mathrm{Acc}_k$ under the standard full-audio setting. We compute Pearson's correlation coefficient $r$ between $\{R_k^{(l^*)}\}$ and $\{\mathrm{Acc}_k\}$ across all tasks\footnote{We exclude three tasks with extremely small sample sizes (e.g., 15) to avoid unstable correlation estimates: \emph{Imagination}, \emph{Aesthetic Evaluation}, and \emph{Audio Difference Analysis}.}. 
\begin{figure}[t]
    \centering
    \includegraphics[width=\linewidth]{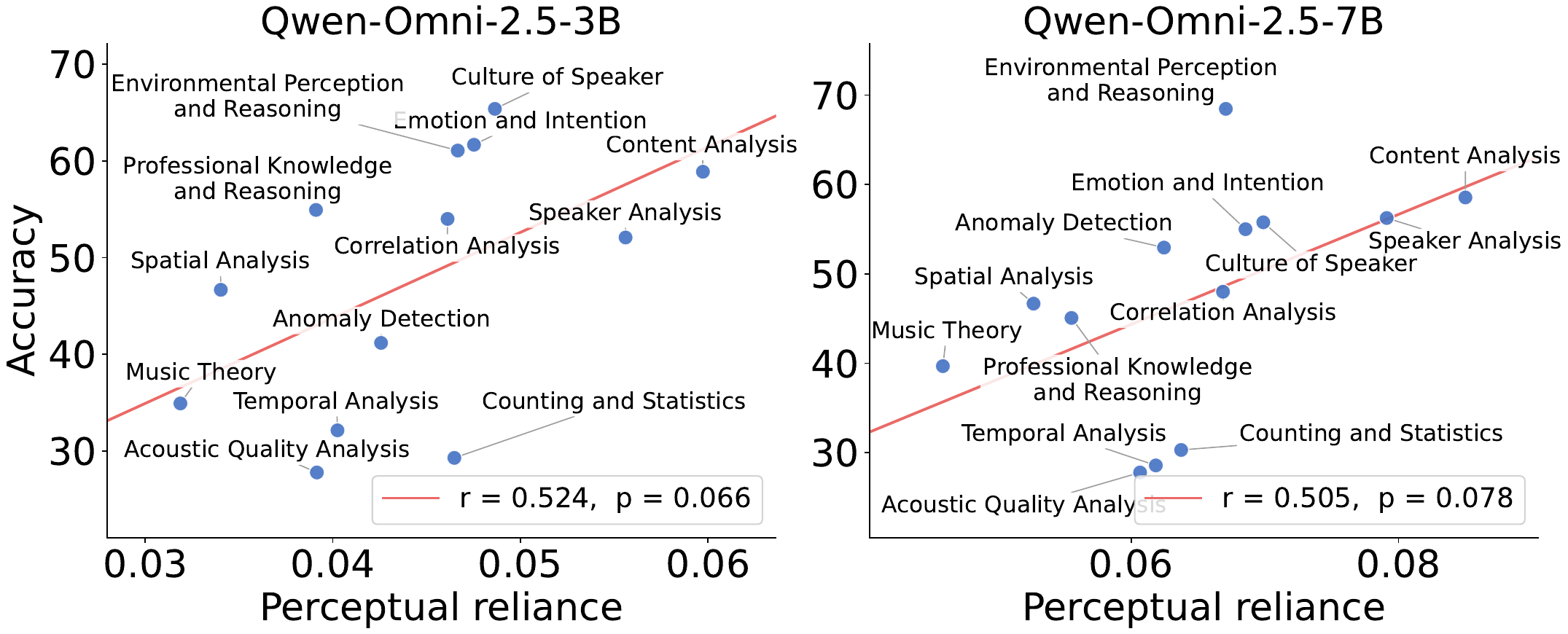}\vspace{-3pt}
    \caption{Relationship between perceptual reliance and accuracy at task-level on MMAR for Qwen2.5-Omni-3B and Qwen2.5-Omni-7B.}
    \vspace{-10pt}
    \label{fig:correlation_3b_7b}
\end{figure}
Figure~\ref{fig:correlation_3b_7b} plots task accuracy against perceptual reliance. Both model scales exhibit positive correlations with moderate significance, with $r=0.524$ for Omni-3B and $r=0.505$ ($p<0.1$) for Omni-7B ($p<0.1$). 
Tasks for which the model integrates more acoustic evidence also tend to achieve higher overall performance. 


Although this result establishes a clear association between perceptual reliance and task performance, it does not reveal whether stronger performance is driven by richer acoustic representations or by greater utilisation of that evidence during subsequent reasoning (RQ3). 

\subsection{Causal Contribution of Acoustic Perception}
To answer \textbf{RQ3}, we 
ask whether acoustic perception is \emph{causally necessary} for successful task
performance. 
We estimate the causal contribution of acoustic perception by applying the same attention-masking intervention. 
For each task, we compare the standard full-audio condition with a perturbed condition: 
\begin{equation}
\Delta_k
=
\mathrm{Acc}_k^{\mathrm{full}}
-
\mathrm{Acc}_k^{\mathrm{mask}},
\end{equation}
A larger $\Delta_k$ indicates that the model relies on acoustic perception, whereas a
small $\Delta_k$ suggests that the task can largely be solved through language-side
reasoning alone.

\begin{figure}[t]
    \centering
    \includegraphics[width=\linewidth]{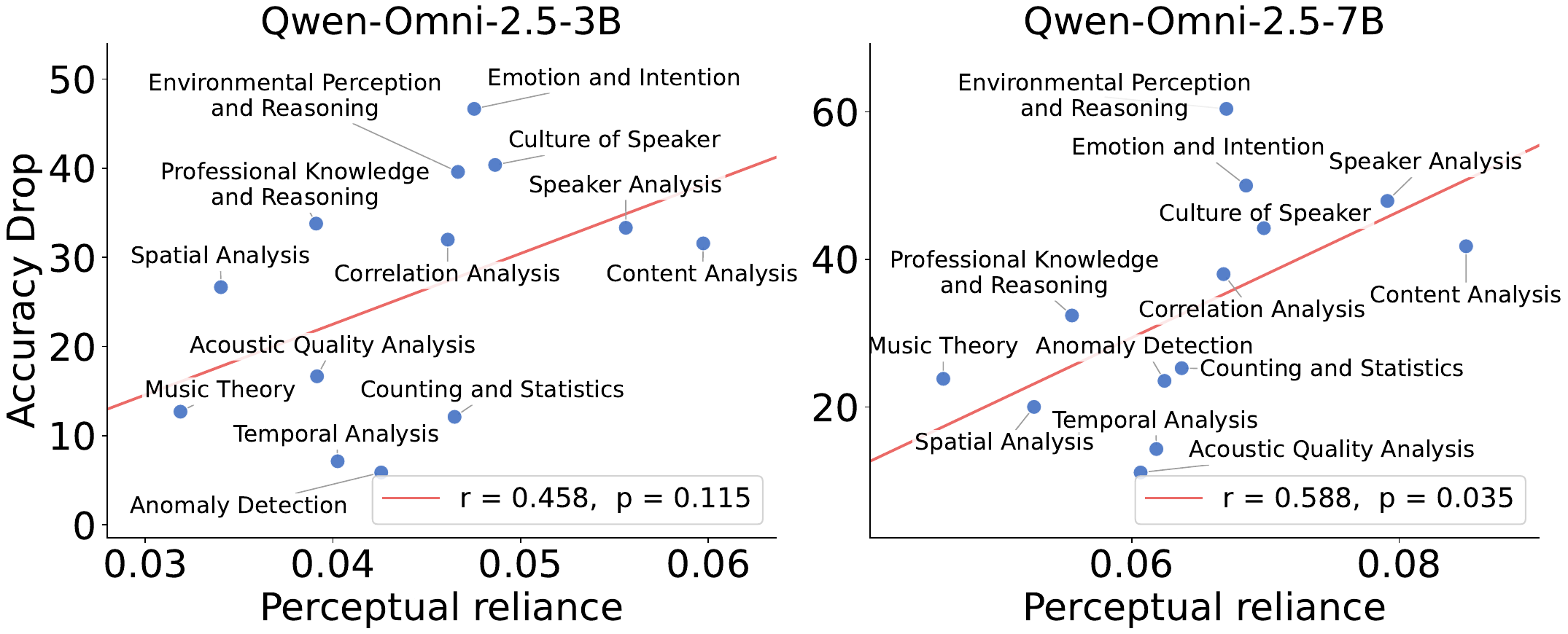}\vspace{-3pt}
    \caption{Relationship between perceptual reliance and the absolute accuracy drop at task-level on MMAR for Qwen2.5-Omni-3B and Qwen2.5-Omni-7B.}
    \vspace{-10pt}
    \label{fig:causual_3b_7b}
\end{figure}

As shown in Figure~\ref{fig:causual_3b_7b}, tasks with higher perceptual reliance generally experience larger performance degradation when acoustic evidence is removed. 
Omni-7B exhibits a moderate positive correlation of $r=0.588$, while Omni-3B shows the same directional pattern with $r=0.458$ between the perception reliance and performance drops. This pattern suggests that stronger perceptual reliance is associated with downstream utilization of acoustic evidence in reasoning, which in turn contributes to better task performance.



\paragraph{Implications for Post-training.}
These findings suggest that perceptual grounding is a desirable optimization target. If successful reasoning depends on effective acoustic grounding, then post-training should explicitly encourage models to strengthen this grounding rather than optimizing only the final prediction. Existing RL-based post-training, however, optimizes answer correctness alone. Such rewards evaluate only the final outcome and are agnostic to whether the prediction was grounded in acoustic evidence or obtained through language-side shortcut reasoning. Consequently, RL has no explicit incentive to improve the perceptual capability that our analysis identifies as critical for many tasks. 
\section{Perception-Grounded TTRL} 
We propose \sysname{}, a perception-grounded TTRL framework that aligns the optimization signal with the acoustic perception, encouraging the model to incorporate acoustic evidence when solving audio tasks. 

\subsection{Problem Setup and Overview} 
\label{sec:method-setup}
As shown in Figure~\ref{fig:framework_overview}, let $\pi_\theta$ denote the policy model and $\pi_{\mathrm{ref}}$ a frozen copy of the 
base model. A prompt $x = (a, q, \{o_1, \dots, o_J\})$ consists of an audio clip $a$, 
a natural language question $q$, and $J$ candidate answer options. Given $x$, the 
model samples $N$ reasoning trajectories $y_1, \dots, y_N \sim \pi_\theta(\cdot \mid x)$, 
where $y_i = (y_{i,1}, \dots, y_{i,T_i})$ is the $i$-th generated token sequence of 
length $T_i$, 
and $\hat{o}_i \in \{o_1,\dots,o_J,\varnothing\}$ 
denotes the final answer parsed from trajectory $y_i$, with $\varnothing$ marking a parse failure. 
In the absence of ground-truth labels, we use the majority answer $m = \mathrm{mode}_{\,i:\,\hat{o}_i \neq \varnothing} \hat{o}_i$ 
as a pseudo-label to provide a 
correctness signal.

Our goal is to update $\pi_\theta$ to improve audio reasoning without ground-truth 
supervision, while ensuring that the optimisation signal is aligned with the causal 
role of acoustic perception. \sysname{} achieves this through three steps: 
(\textit{i}) computing a trajectory-level perceptual grounding score that quantifies 
how much each trajectory relies on acoustic evidence; (\textit{ii}) deriving a 
reliability weight that reflects how strongly acoustic grounding 
distinguishes the majority answer from alternatives; and (\textit{iii}) incorporating 
this reliability weight into a tempered advantage for policy optimization. 

\subsection{Trajectory-level Perceptual Grounding Score} 
\label{sec:method-grounding}

We first quantify how much each trajectory \(y_i\) relies on acoustic evidence, aiming to assign a higher score for greater reliance. We evaluate its generated tokens under two teacher-forcing passes: the original condition and the masked condition, in which textual tokens are prevented from attending to audio tokens.
For each predicted token $y_{i,t}$, we compute the \emph{per-token audio dependence} as the log-probability gap between the two conditions:
\begin{equation}
\Delta_{i,t}
=
\log \pi_{\theta,\mathrm{full}}
\!\left(y_{i,t}\mid a,\,y_{i,<t}\right)
-
\log \pi_{\theta,\mathrm{mask}}
\!\left(y_{i,t}\mid a,\,y_{i,<t}\right).
\end{equation}
A large positive $\Delta_{i,t}$ indicates that the token is informed by the acoustic content rather than language priors alone. Aggregating over all tokens in the trajectory, we define the \emph{perceptual grounding score} of trajectory $y_i$ as the mean 
log-probability gap:
\begin{equation}
g_i
=
\frac{1}{T_i}
\sum_{t=1}^{T_i}
\Delta_{i,t}.
\end{equation}
A larger $g_i$ indicates that the reasoning trajectory relies more heavily on acoustic 
evidence. 

\subsection{Reliability Weight}
\label{sec:method-support}
The grounding score $g_i$ alone is insufficient to 
shape the update, as it reflects only the degree of acoustic reliance and not the correctness of the resulting reasoning. A trajectory may exhibit high $g_i$ while arriving at an incorrect answer, and naively incorporating $g_i$ into the 
advantage would risk reinforcing well-grounded but incorrect trajectories.

To address this, we condition the perceptual signal on the pseudo-label $m$ by 
measuring whether trajectories that support the majority answer are more acoustically grounded than those that dissent. Formally, we define the \emph{acoustic grounding margin} as:
\begin{equation}
    s = \underbrace{\operatorname{mean}_{\,i:\,\hat{o}_i = m} g_i}_{\text{majority}}
      - \underbrace{\operatorname{mean}_{\,i:\,\hat{o}_i \notin \{m,\varnothing\}} g_i}_{\text{dissent}}.
    \label{eq:agm}
\end{equation}
A large positive $s$ indicates that trajectories supporting the majority answer rely 
more strongly on acoustic evidence than dissenting trajectories, lending perceptual 
credibility to the pseudo-label. When $s \approx 0$, acoustic grounding does not 
distinguish between the two groups, and the pseudo-label receives little perceptual 
support. We therefore interpret $s$ as a direct measure of how strongly acoustic 
evidence corroborates the majority vote.

To obtain a bounded reliability weight, we normalize $s$ 
within each minibatch and map it monotonically to $r \in [r_{\min}, r_{\max}]$, such that a 
larger acoustic grounding margin yields a larger reliability weight $r$.



\subsection{Policy Optimization} 
\label{sec:method-advantage}

The proposed 
\sysname{} replaces the standard GRPO advantage with 
a reliability-weighted vote margin, ensuring that the optimisation signal is amplified when acoustic evidence supports the majority vote and attenuated when it does not.

\paragraph{Perception-Grounded Advantage.}
Standard GRPO optimizes a clipped surrogate objective over a group of rollouts, 
regularized by a KL penalty to a frozen reference model, where the advantage of each rollout reflects its relative quality within the group. In \sysname{}, 
we define the group-relative advantage with a perception-grounded one that scales the vote margin by the acoustic reliability of the pseudo-label. Let \(v(\hat{o}_i)\) denote the fraction of trajectories supporting answer \(\hat{o}_i\), and let \(1/J\) denote the uniform vote share across the \(J\) candidate answers. We refer to
\(v(\hat{o}_i)-1/J\) as the \emph{vote margin}, which measures how strongly the sampled trajectories favour \(\hat{o}_i\) relative to a uniform distribution. The perception-grounded advantage is:
\begin{equation}
    A_i = r \cdot \left(v(\hat{o}_i) - \frac{1}{J}\right),
    \label{eq:advantage}
\end{equation}
When $s$ is largely 
and $r \to r_{\max}$, the vote margin drives the update most strongly; when $s$ is small 
and $r \to r_{\min}$, the vote margin is down-weighted to the floor $r_{\min}$ without being fully discarded, since $r_{\min}>0$. 
The fixed baseline $1/J$, rather than the group mean, ensures that a prompt on which all rollouts agree still carries a positive training signal $r \times (1 - 1/J)$. The full procedure is summarized in Algorithm~\ref{alg:agrt}.

\paragraph{Objective.}
Substituting $A_i$ into the GRPO objective, the full perception-grounded loss is:
\begin{equation}
\begin{split}
    \mathcal{L} = 
    &-\,\mathbb{E}\!\left[\min\!\left(\rho_{i,t} A_i,\;
    \operatorname{clip}(\rho_{i,t}, 1{-}\epsilon_c, 1{+}\epsilon_c)\, 
    A_i\right)\right] \\
    &+ \beta\, \mathrm{KL}_{k3}\!\left(\pi_\theta \,\|\, 
    \pi_{\mathrm{ref}}\right),
\end{split}
\end{equation}
where $\rho_{i,t} = \pi_\theta(y_{i,t} \mid \cdot) / \pi_{\mathrm{old}}(y_{i,t} \mid 
\cdot)$ is the importance sampling ratio. The perception-grounded signal enters entirely through $A_i$, no additional loss term or reward scalar is introduced, keeping the framework minimal and directly compatible with standard GRPO implementations. 

\begin{algorithm}[t]
\caption{\sysname{} for one minibatch}
\label{alg:agrt}
\begin{algorithmic}[1]
\Require prompts $\{x_b\}$, policy $\pi_\theta$, running scale state
\For{each prompt $x_b$}
  \State sample $N$ rollouts; parse letters $\hat o_i$; form vote $v$, majority $m$
  \State two forward passes (original audio input / perturbed condition) $\Rightarrow \Delta_{i,t} \Rightarrow g_i$
  \State $s_b \gets \operatorname{mean}_{i:\hat o_i=m} g_i - \operatorname{mean}_{i:\hat o_i\notin\{m,\varnothing\}} g_i$
\EndFor
\State standardize defined $\{s_b\}$ by minibatch mean and cumulative scale $\Rightarrow r_b$ (clip to $[r_{min},r_{max}]$)
\For{each prompt, each rollout $i$}
  \State if $\hat o_i=\varnothing$: $A_i \gets 0$;\quad else $A_i \gets r_b\,(v(\hat o_i)-1/J)$, then if $A_i<0$: $A_i \gets \gamma A_i$
\EndFor
\State drop groups with $\max_i|A_i|\le\epsilon$; return $\{A_i\}$ for the GRPO update
\end{algorithmic}
\end{algorithm}

\section{Experimental Setup}
\label{sec:exp}

\paragraph{Datasets. }
\label{sec:exp-data}
We evaluate on the test sets of two widely used multiple-choice audio reasoning benchmarks. \textbf{MMAR}~\cite{ma2026mmar} contains $1{,}000$ examples across speech, general audio, music, and mixed-audio settings, organized into $16$ sub-categories and four reasoning levels: Signal, Perception, Semantic, and Cultural. \textbf{MMAU}~\cite{sakshi2025mmau} contains $1{,}000$ examples from its \emph{test-mini} split across speech, sound, and music, organized into $27$ sub-categories. For both benchmarks, the prompts are used for TTRL without accessing the ground-truth answers, which are reserved exclusively for evaluation.

\paragraph{Models and Baselines. }
\label{sec:exp-models}
We apply \sysname{} to two LALMs: Qwen2.5-Omni-3B and Qwen2.5-Omni-7B~\cite{xu2025qwen25omni}. 
For all models, we freeze the original parameters and optimise only Low-Rank Adaptation (LoRA) modules~\cite{hu2022lora}. We optimise rank $r=8$ and scaling factor $\alpha=16$, with no dropout, and insert the adapters into the linear projection layers of the model, run on one H100 and one A100 GPU. 

We compare \sysname{} against two baselines. \textbf{Base} denotes the original model without any test-time parameter updates. TTRL~\cite{zuo2026ttrl} shares the same GRPO optimisation as \sysname{}, but uses the binary majority-vote reward $\mathbf{1}[\hat{o}_i=m]$, 
with no grounding-aware reliability weighting.

\paragraph{Training Configuration. }
\label{sec:exp-train}
For each prompt we draw $N=8$ rollouts at sampling temperature $T=1.0$ and nucleus $\text{top-}p=0.99$; optimizer steps use a prompt batch of $B=8$
($B\!\times\!N=64$ trajectories per step). We train for a single epoch over the
prompt pool. The learning rate is $3\times10^{-6}$ with a constant schedule;
the KL penalty uses coefficient $\beta=0.05$ with the low-variance (k3) estimator. \sysname{}'s own settings are as follows: reliability bounds $[r_{\min},r_{\max}]=[0.2,0.8]$, negative-advantage damping $\gamma=0.4$,
the running scale taken as the
cumulative average of the per-minibatch support spread, and the per-minibatch standard
deviation computed with the unbiased ($n{-}1$) estimator.

Three additional design choices ensure stable training. Groups whose advantages are 
all within $\epsilon$ of zero, indicating either near-uniform vote distributions or 
an absence of any parseable answer are dropped from the update, as they carry no 
reliable learning signal. Negative advantages are damped by $\gamma$, slowing 
the suppression of minority answers and reducing the risk of premature collapse to a 
single answer. Trajectories that fail to parse are assigned $A_i = 0$. Evaluation seeds are derived by hashing the global seed 42 with each prompt ID, while training uses the global seed directly.

\paragraph{Evaluation Protocol. }
\label{sec:exp-eval}

We report four complementary metrics. \emph{Greedy} evaluates the model's deterministic prediction using a single completion. 
For sampling-based evaluation, we draw 8 trajectories, 
and evaluate \emph{Maj@8} which measures whether the majority answer among these eight samples is correct, 
and \emph{Avg@8} that reports the mean accuracy across the eight individual samples. 
\emph{Pass@$k$} estimates the probability that at least one correct solution is found among $k$ sampled outputs, reflecting the model’s ability to produce a correct answer under different sampling budgets. 

\section{Results}
\subsection{Effectiveness of PG--TTRL} 

\begin{table*}[t]
\centering
{\small
\setlength{\tabcolsep}{10pt}
\renewcommand{\arraystretch}{1.1}

\begin{tabular}{llcccccc}
\toprule
\multirow{2}{*}{Model}
& \multirow{2}{*}{Method}
& \multicolumn{3}{c}{MMAR}
& \multicolumn{3}{c}{MMAU} \\
\cmidrule(lr){3-5}
\cmidrule(lr){6-8}
&
& Greedy $\uparrow$
& Avg@8 $\uparrow$
& Maj@8 $\uparrow$
& Greedy $\uparrow$
& Avg@8 $\uparrow$
& Maj@8 $\uparrow$ \\
\midrule

\multirow{3}{*}{Qwen2.5-Omni-3B}
& Base
& 53.1
& 52.2
& 56.7
& 69.9
& 67.9
& 74.0 \\

& TTRL
& 53.4
& 52.4
& 56.5
& \textbf{70.7}
& 68.7
& \textbf{74.4} \\

& PG-TTRL
& \textbf{55.7}
& \textbf{53.8}
& \textbf{58.2}
& 70.5
& \textbf{69.3}
& 74.1 \\
\midrule

\multirow{3}{*}{Qwen2.5-Omni-7B}
& Base
& 53.2
& 47.3
& \textbf{61.6}
& 69.8
& 62.3
& 75.1 \\

& TTRL
& 55.9
& 48.0
& 61.4
& 70.7
& 64.2
& \textbf{75.4} \\

& PG-TTRL
& \textbf{56.0}
& \textbf{51.9}
& 61.4
& \textbf{71.3}
& \textbf{67.2}
& 74.6 \\

\bottomrule
\end{tabular}
}\vspace{-3pt}
\caption{
Results (Greedy, Avg@8, Maj@8) on MMAR and MMAU.
The best result within each model group is shown in \textbf{bold}.
}
\label{tab:main-results}
\end{table*}
Table~\ref{tab:main-results} 
summarises the main results on MMAR and MMAU. Across both benchmarks and both model scales, \sysname{} achieves the strongest Greedy and Avg@8 results, with the largest gains reaching $4.6$ points on MMAR and $4.9$ points on MMAU. This demonstrates that perception-grounded optimisation consistently improves both deterministic and expected reasoning performance. The gains are particularly pronounced on MMAR, which contains tasks that are more complex and auditory in nature and therefore benefit more directly from stronger acoustic grounding.

The improvement in Avg@8 consistently exceeds that in Greedy across both models, most strikingly for the 7B model on MMAR, where Avg@8 improves by $+4.6$ points against a Greedy gain of $+2.8$ points. This suggests that \sysname{} broadens the set of trajectories capable of reaching the correct answer rather than 
simply sharpening the single most likely prediction, consistent with the perception-grounded advantage \emph{discouraging over-commitment to a single answer pathway and preserving diverse but acoustically supported reasoning trajectories.}

While \sysname{} improves Greedy and Avg@8 consistently, Maj@8 gains are smaller and 
in some cases comparable to Base or TTRL. This suggests that \sysname{} raises per-trajectory accuracy without necessarily concentrating probability mass on a single dominant answer. In the label-free setting, 
the model is not explicitly trained to maximise majority agreement, but rather to ground each trajectory in acoustic evidence, which may distribute probability more 
evenly across correct and near-correct answers.

Both the 3B and 7B models benefit from \sysname{}, indicating that perception-grounded 
optimisation is not specific to a particular model capacity. The 7B model shows a 
notably larger Avg@8 improvement on MMAU ($+4.9$ over Base), suggesting that larger 
models may have greater latent perceptual capacity that \sysname{} is better able to 
unlock.

\subsection{Performance under different sampling budgets}
\begin{figure}[t]
    \centering
    \vspace{-10pt}
    \includegraphics[width=\linewidth]{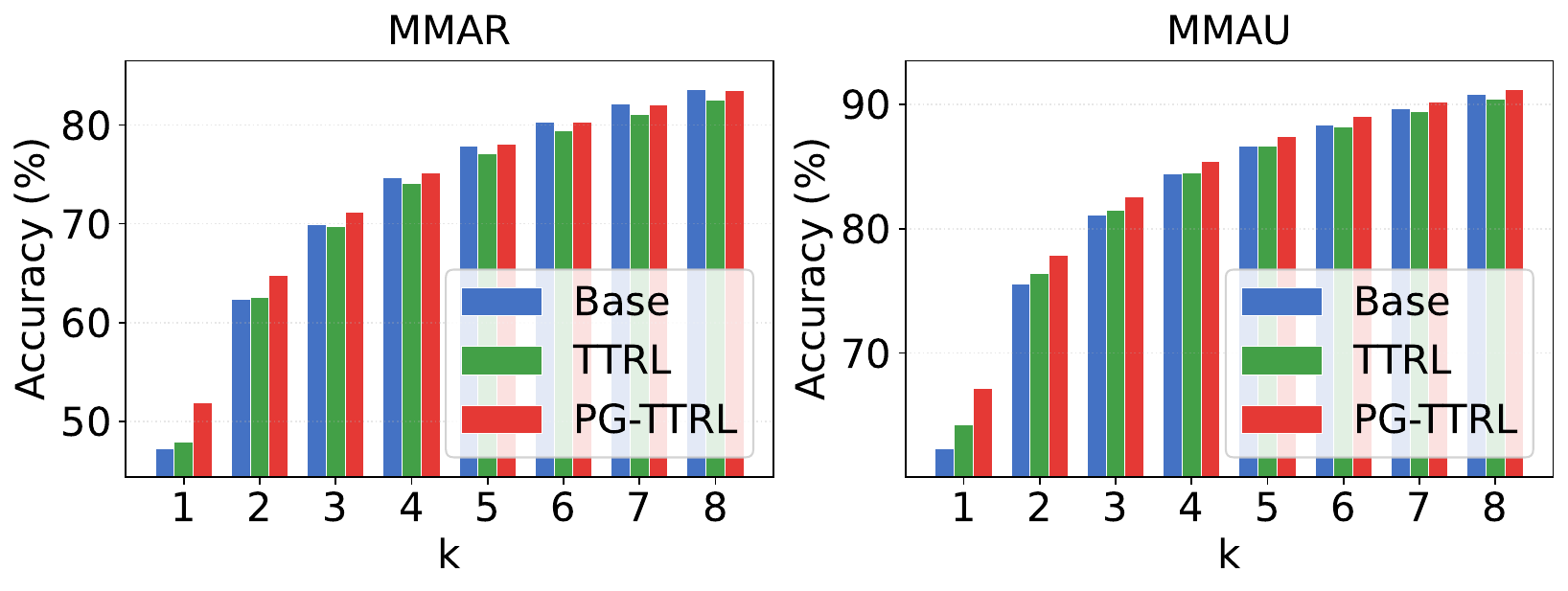}
    \vspace{-20pt}
    \caption{Pass@k (k = 1…8) comparison on MMAR and MMAU for the Qwen-Omni-2.5-7B model.}
    \vspace{-10pt}
    \label{fig:perceptual_reliance}
\end{figure}

We further evaluate Pass@$k$ to assess performance under different test-time sampling 
budgets. \sysname{} achieves the highest Pass@$k$ across nearly all values of $k$, with \emph{the largest gains at small $k$}, particularly significant in practice, where limited inference budgets make early correct solutions most valuable.

Unlike TTRL, whose Pass@$k$ falls below the base model at larger $k$, \sysname{} remains consistently stronger across the full range. For difficult instances, the base 
model may reach the correct answer only through low-probability trajectories, leading to an incorrect majority pseudo-label. TTRL then concentrates probability mass on the wrong answer and suppresses these rare correct paths, a self-reinforcing failure that worsens with more sampling. In contrast, \sysname{}'s perception-grounded advantage 
discounts updates lacking acoustic support, preventing over-commitment to an incorrect 
majority vote and preserving low-probability correct trajectories across all sampling 
budgets.

\subsection{What Does Post-Training Actually Improve?}

\begin{figure}[t]
    \centering
    \vspace{-5pt}
    \includegraphics[width=\linewidth]{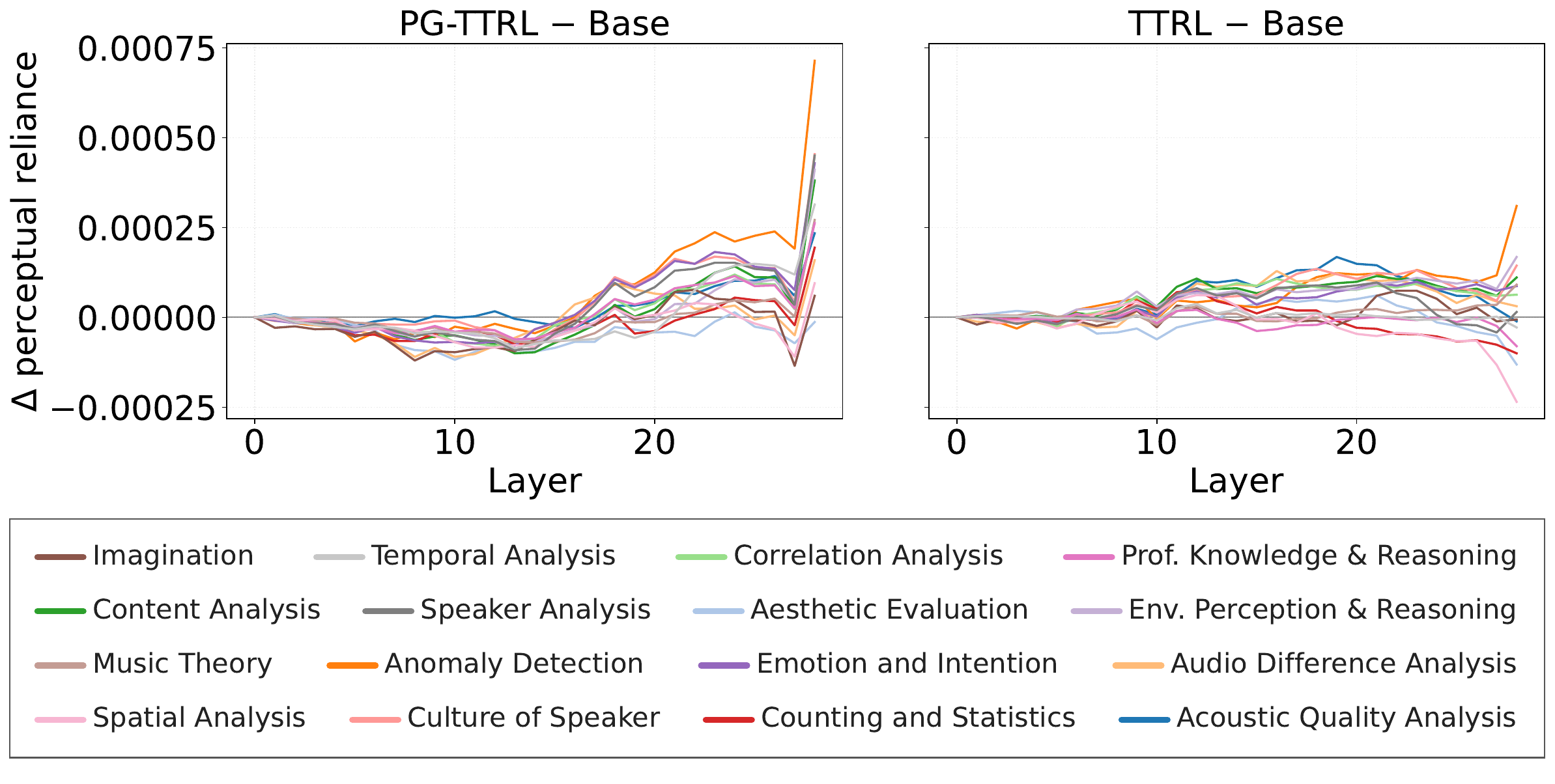}
    \vspace{-17pt}
    \caption{Layer-wise changes in perceptual reliance relative to the base Qwen2.5-Omni-7B model on MMAR.}
    \vspace{-10pt}
    \label{fig:reliance_change}
\end{figure}

Figure~\ref{fig:reliance_change} compares the layer-wise changes in perceptual 
reliance produced by \sysname{} and TTRL relative to the base model on MMAR for 
Qwen2.5-Omni-7B. TTRL introduces a modest positive shift from relatively early layers, 
where cross-modal transfer is largely completed. In contrast, \sysname{} produces 
substantially larger gains in the later backbone layers, indicating that it primarily 
reinforces acoustic utilisation during late-stage reasoning rather than reshaping 
earlier perceptual integration. 
This pattern is consistent with PG-TTRL design. Its reward signal reinforces trajectories that draw on acoustic evidence, and gradient signals propagating from the output naturally have the strongest effect on layers closest to the output, with increasing attenuation toward earlier layers. More importantly, early-layer audio representations are largely determined by the audio encoder, projection layer, and pretraining, leaving limited room for post-training reward signals to reshape how audio information is initially captured and transferred across modalities. Overall, these results suggest that PG-TTRL successfully improves audio utilization during late-stage reasoning. Future RL-based post-training may benefit from co-training with a more capable audio encoder or stronger multimodal alignment to address the perception bottleneck in earlier layers. 


\section{Conclusion}
We investigated how audio information is integrated and utilised throughout LALM reasoning, and showed that stronger perceptual reliance is associated with better task performance and a greater causal contribution from the audio input. Building on these findings, we proposed PG-TTRL, which optimizes the model to reason based on perception at test time without ground-truth labels. Experiments across multiple models and benchmarks demonstrate consistent improvements in audio reasoning. A current limitation is that the perceptual grounding score is computed only from generated CoT tokens. Future work could incorporate hidden representations to 
provide a richer and more direct signal of acoustic grounding for optimisation.
\bibliography{ours}


\end{document}